\documentclass{article}
\usepackage[utf8]{inputenc}
\usepackage{amsthm}
\usepackage{graphicx,ae,aecompl}
\usepackage{amssymb}
\usepackage{amsmath}
\usepackage{subfigure}
\usepackage{algorithmic}
\usepackage[boxruled,linesnumbered]{algorithm2e}
\usepackage{makeidx}
\usepackage{float}
\usepackage{scalefnt}
\usepackage{indentfirst}

\title{A wavelet-based method in aggregated functional data analysis}
\author{Alex Rodrigo dos S. Sousa \\ University of Campinas, Brazil \\ }

\begin{document}

\numberwithin{equation}{section}
\numberwithin{table}{section}
\numberwithin{figure}{section}

 \maketitle
    \begin{abstract}
       In this paper we consider aggregated functional data composed by a linear combination of component curves and the problem of estimating these component curves.  We propose the application of a bayesian wavelet shrinkage rule based on a mixture of a point mass function at zero and the logistic distribution as prior to wavelet coefficients to estimate mean curves of components. This procedure has the advantage of estimating component functions with important local characteristics such as discontinuities, spikes and oscillations for example, due the features of wavelet basis expansion of functions. Simulation studies were done to evaluate the performance of the proposed method and its results are compared with a spline-based method. An application on the so called tecator dataset is also provided. 
        
    \end{abstract}

\section{Introduction}
The statistical problem of estimating individual mean curves from aggregated samples composed by linear combinations of these curves (with known coefficients) plus a Gaussian random error appears in several areas of science. In spectroscopy for example, there is an interest in estimating individual mean absorbance curves of the compositions of a given substance from samples of absorbance curves of the substance itself. In this case, by the Beer-Lambert Law (Brereton, 2003), the absorbance of the substance is formed by the linear (convex) combination of the absorbances of its constituents and the coefficients of the linear combinations are the concentrations of the constituents in the substance. This problem is commonly called calibration in chemometrics. Another example appears in the composition of electricity consumption in a given region, in which the energy consumption curve for a period of time is composed of the consumption curves of each consumer. For more details on these examples, see Dias et al. (2013).

Several statistical methods have been proposed to estimate individual mean curves from aggregated curves. Many of these methods consider the data structure as multivariate due to the fact that observations are obtained at discrete times, so that each point of the aggregated curve in the sample is considered as a variable with a given correlation structure with the other points. Principal Components Regression (PCR) by Cowe and McNicol (1985), Partial Least Squares (PLS) by Wold et al. (1983) are important proposed multivariate methodologies to deal with aggregated data. Bayesian approaches and wavelet-based methods were also proposed by Brown et al. (1998a,b) and Brown (2001). Although being successfully applied in many cases, such methods fail to consider the functional structure of the data, despite the limitation of discrete measurements due to instrumentation. In this sense, well succeeded estimation proposals using the functional approach were made by Dias et al. (2009) and Dias et al. (2013) using expansion of the curves by splines or B-splines. In this case, although the expansions by splines have excellent performance in smooth curves, such performances decrease when the individual curves present local characteristics, such as discontinuities, peaks or oscillations, for example. For a great overview about functional data analysis and basis expansion of functions, see Ramsay and Silverman (2005).

In this way, we propose in this work the use of wavelet basis to estimate individual mean curves from aggregated curves. In addition, we propose the application of the bayesian wavelet shrinkage rule based on a mixture of a point mass function at zero and the logistic distribution as prior to the wavelet coefficients, proposed by Sousa (2020) to estimate the wavelet coefficients from empirical wavelet coefficients obtained by the application of a discrete wavelet transformation on the original aggregated dataset. 

Wavelet expansions have characteristics that make them quite useful from the point of view of function representation since they are located in both time and scale in an adaptive way. Their coefficients are typically sparse, they can be obtained by fast
computational algorithms, and the magnitudes of coefficients are linked with the smoothness properties of the functions they represent. These properties enable time/frequency data analysis, bring computational advantages, and allow for statistical
data modeling at different resolution scales.

Wavelet shrinkage methods are used to estimate the coefficients associated with the representation of the function in the wavelet domain by reducing the magnitude of the observed (empirical) coefficients that are obtained by the wavelet transform in the the original data. There are in fact several shrinkage techniques available in the literature. The main works in this area are of Donoho e Johnstone (1994, 1995), but also Donoho et al. (1995, 1996a, 1996b) can be cited. For more details of wavelets and wavelet shrinkage methods, see Daubechies (1992), Mallat (1998), Vidakovic (1999) and Jansen (2001).

In this context, bayesian wavelet shrinkage methods have been extremely studied and applied in denoising problems in many statistical areas and science fields. In curve estimation problem, the focus of this work, these methods are interesting due the possibility to incorporate prior information about the wavelet coefficients of the wavelet transformed unknown function, such as sparsity, support (if they are bounded), local features, and others. Further, the use of the logistic distribution is very interesting in the context of wavelet coefficients estimation, once it is symmetric (around zero in our proposal) and its scale parameter is easily related with the degree of shrinkage of the associated bayesian rule, which can be usefull to elicite it in practical situations, see Sousa (2020).

This paper is organized as follows: Section 2 defines the statistical model involving the aggregated functional curve as linear combination of curves plus a gaussian random error. The estimation procedure based on wavelets expansion and wavelet shrinkage is described in Section 3. Simulation studies to evaluate the performance of the proposed methodology in terms of averaged mean squared error and to compare it with a methodology based on splines expansion are done in Section 4. An application of the proposed procedure on the so called Tecator dataset is shown in Section 5. Conclusion and further considerations are in Section 6.

\section{Statistical Model}

We consider a univariate function $A(t) \in \mathbb{L}_2(\mathbb{R})=\{f: \mathbb{R} \rightarrow \mathbb{R}|\int f^2 < \infty\}$ that can be written as
\begin{equation}\label{funmodel}
A(t) = \sum_{l=1}^{L}y_{l} \alpha_{l}(t) + e(t),
\end{equation}
where $\alpha_{l}(t) \in \mathbb{L}_2(\mathbb{R})$ are unknown component functions, $y_{l}$ are known real valued weights, $l=1, \cdots, L$, and $\{e(t),t \in \mathbb{R}\}$ is a zero mean gaussian process with unknown variance $\sigma^2$, $\sigma > 0$. The statistical problem here concerns the estimation of the functions $\alpha_l(t)$. To do so, we expand each function $\alpha_l$ of \eqref{funmodel} in wavelet basis as
\begin{equation} \label{alphawav}
\alpha_l(t) = \sum_{j,k \in \mathbb{Z}} \gamma_{jk}^{(l)} \psi_{jk}(t), \hspace{1cm} l=1, \cdots, L,
\end{equation}
where $\{\psi_{jk}(x) = 2^{j/2} \psi(2^j x - k),j,k \in \mathbb{Z} \}$ is an orthonormal wavelet basis for $\mathbb{L}_2(\mathbb{R})$ constructed by dilations $j$ and translations $k$ of a function $\psi$ called wavelet or mother wavelet and $\gamma_{jk}^{(l)}$ 's are unknown wavelet coefficients of the expansion of the component function $\alpha_l$. Note that we consider the same wavelet family for all component functions expansion. In this sense, the problem of estimating the functions $\alpha_l$ becomes a problem of estimating the finite number of wavelet coefficients $\gamma_{jk}^{(l)}$'s of the representation \eqref{alphawav}.

In practice, one observes $I$ samples of the aggregated curve $A(t)$ at $M = 2^J$ locations $t_1, \cdots, t_M$, i.e, our dataset is $\{(t_m, A_i(t_m))$, $m = 1, \cdots, M $ and $i = 1, \cdots, I\}$. Thus, the discrete version of \eqref{funmodel} is 
\begin{equation}\label{discmodel}
A_i(t_m) = \sum_{l=1}^{L}y_{il}\alpha_l(t_m) + e_i(t_m), \hspace{0.5cm} i=1,\cdots,I, \hspace{0.5cm} m=1,\cdots,M=2^J,
\end{equation}
where $e_i(t_m)$ are independent and identically normal distributed random noise with zero mean and variance $\sigma ^2$, $\forall i,m$. Further, the $I$ samples are obtained at the same locations $t_1, \cdots, t_M$ but the weights of the linear combinations are allowed to be different from one sample to another. We can rewrite \eqref{discmodel} in matrix notation as
\begin{equation}\label{vecmodel}
\boldsymbol{A} = \boldsymbol{\alpha}\boldsymbol{y} + \boldsymbol{e},
\end{equation}
where $\boldsymbol{A} = (A_{mi}=A_i(t_m))_{1 \leq m \leq M, 1 \leq i \leq I}$, $\boldsymbol{\alpha} = (\alpha_{ml}=\alpha_l(t_m))_{1 \leq m \leq M, 1 \leq l \leq L}$, $\boldsymbol{y} = (y_{li})_{1 \leq l \leq L, 1 \leq i \leq I}$ and $\boldsymbol{e} = (e_{mi}=e_i(t_m))_{1 \leq m \leq M, 1 \leq i \leq I}$.

The wavelet shrinkage procedure will be made in the wavelet domain to estimate the coefficients $\gamma$'s of \eqref{alphawav}. For this reason, we apply a discrete wavelet transform (DWT) on the original aggregated data, which can be represented by a $M \times M$ wavelet transformation matrix $\boldsymbol{W}$, and applied on both sides of \eqref{vecmodel}, i.e,
\begin{align}\label{matrixmodel}
\boldsymbol{W}\boldsymbol{A} &= \boldsymbol{W}(\boldsymbol{\alpha}\boldsymbol{y} + \boldsymbol{e}) \nonumber \\
\boldsymbol{W}\boldsymbol{A} &= \boldsymbol{W}\boldsymbol{\alpha}\boldsymbol{y} + \boldsymbol{W}\boldsymbol{e} \nonumber \\
\boldsymbol{D} &= \boldsymbol{\Gamma}\boldsymbol{y} + \boldsymbol{\varepsilon},
\end{align}
where $\boldsymbol{D} = \boldsymbol{W}\boldsymbol{A} = (d_{mi})_{1 \leq m \leq M, 1 \leq i \leq I}$ is the matrix with the empirical wavelet coefficients of the aggregated curves, $\boldsymbol{\Gamma} = \boldsymbol{W}\boldsymbol{\alpha} = (\gamma_{ml})_{1 \leq m \leq M, 1 \leq l \leq L}$ is the matrix with the unknown wavelet coefficients of the component curves and $\boldsymbol{\varepsilon} = \boldsymbol{W}\boldsymbol{e} =  (\varepsilon_{mi})_{1 \leq m \leq M, 1 \leq i \leq I}$ is the matrix with the random errors on the wavelet domain, which remain zero mean normal distributed with variance $\sigma^2$ due the orthogonality property of wavelet transforms. Thus, for a particular empirical wavelet coefficient $d_{mi}$ of $\boldsymbol{D}$, one has the additive model
\begin{equation} \label{coefmodel} 
d_{mi} = \sum_{l=1}^{L} y_{li}\gamma_{ml} + \varepsilon_{mi} = \theta_{mi} + \varepsilon_{mi},
\end{equation}
where $\theta_{mi} = \sum_{l=1}^{L} y_{li} \gamma_{ml}$ and $\varepsilon_{mi}$ is zero mean normal with variance $\sigma^2$, i.e, a single empirical wavelet coefficient of the aggregated curve is also a linear combination of the unknown wavelet coefficients of the component curves plus a random error. Moreover, the weights of this linear combination on wavelet domain remain the same that the original combination of the curves at time domain. 

\begin{figure}[H]
\centering
\subfigure[Bumps function.]{
\includegraphics[scale=0.35]{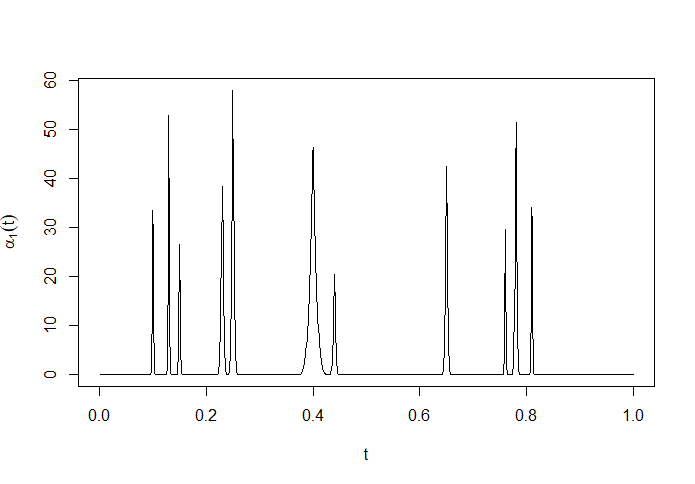}}
\subfigure[Blocks function.]{
\includegraphics[scale=0.35]{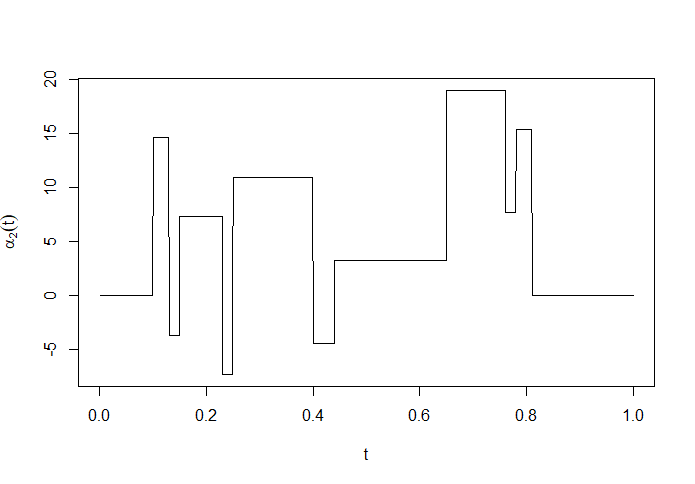}}
\caption{Bumps and blocks functions.} \label{fig:ex1a}
\end{figure}

\begin{figure}[H]
\centering
\subfigure[Generated samples.]{
\includegraphics[scale=0.5]{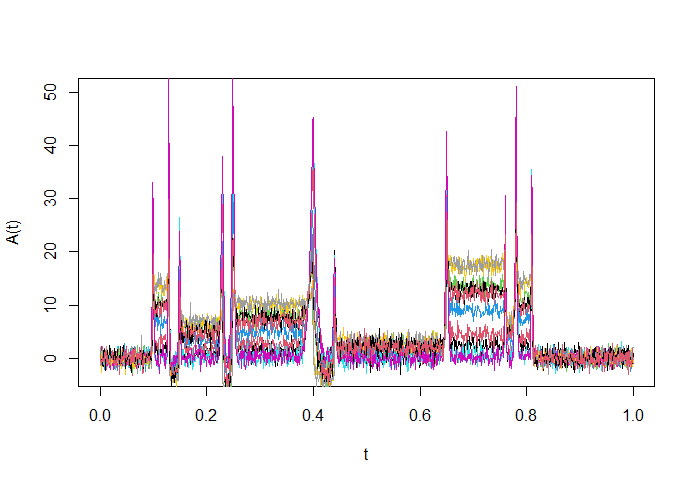}}
\subfigure[Empirical wavelet coefficients.]{
\includegraphics[scale=0.5]{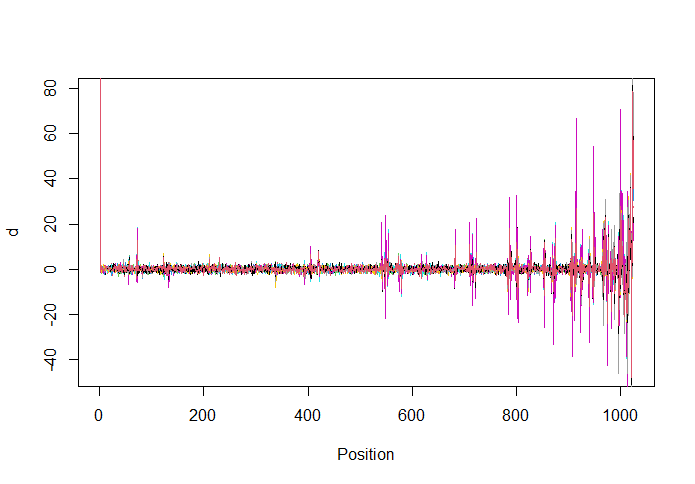}}
\caption{Ten generated samples for mixtures of Bumps and Blocks test functions at 1024 equally spaced locations in $[0,1]$ (a) and their 1024 associated empirical wavelet coefficients (b).} \label{fig:ex1b}
\end{figure}

To illustrate the described data structure and its DWT application, ten samples ($I=10$) of different mixtures of Donoho-Johnstone test functions called Bumps and Blocks ($L=2$), which are shown in Figures \ref{fig:ex1a}(a) and (b), were generated at $M = 1024 = 2^{10}$ equally spaced locations in $[0,1]$. The generated data are displayed in Figure \eqref{fig:ex1b}(a). The empirical wavelet coefficients were obtained by the application of a DWT on the aggregated samples and are shown in Figure \ref{fig:ex1b}(b). 
The two component functions Bumps and Blocks are interesting in this context because they have local features that need to be detected by the estimation procedure. In fact, Bumps has spikes at several locations and Blocks has discontinuities points. The functional expression of these functions are described in Section 4 with other considered test functions in this work. 

Due to the local characteristics of Bumps and Blocks, the generated data by linear combinations have  joint features of them, but in different magnitudes according to the weights of each of the component function. When the weight of the Blocks are greater than the Bumps, the magnitude of the Bumps spikes are reduced for example. But in general, it is possible to visualize Bumps and Blocks characteristics in all the aggregated samples. Moreover, the empirical wavelet coefficients have great magnitudes at the locations of Bumps and Blocks features and are close to zero at locations where these component functions are smooth. This behaviour of the coefficients are typical of wavelet transform. The goal is to recover the component Bumps and Blocks function by the aggregated data composed by linear combinations of them.

\section{Wavelet Shrinkage and Estimation Procedure}
\begin{figure}[H]
\centering
\subfigure[Shrinkage rules.]{
\includegraphics[scale=0.45]{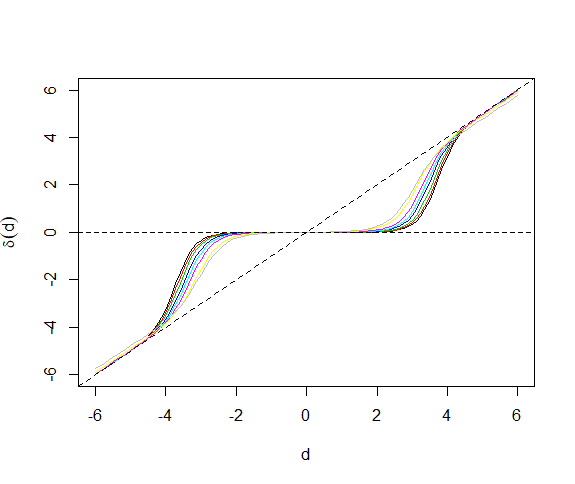}}
\subfigure[Daubechies wavelet function $\psi$ - $N = 10$.]{
\includegraphics[scale=0.45]{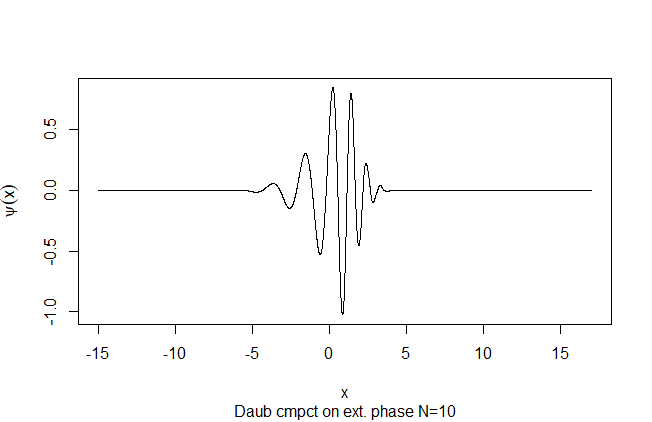}}
\caption{Shrinkage rules under logistic prior obtained numerically for some values of $\tau$ in the interval $(0.5,50)$, $p = 0.9$ and $\sigma = 1$ (a) and Daubechies wavelet function $\psi$ with 10 null moments ($N=10$). }\label{fig:rules}
\end{figure}

The estimation of the wavelet coefficients matrix $\boldsymbol{\Gamma}$  in \eqref{matrixmodel} is done by applying a wavelet shrinkage rule $\delta$ on each single empirical wavelet coefficient $d$, obtaining the matrix $\boldsymbol{\delta(\boldsymbol{D})}$ such that
\begin{equation}
\boldsymbol{\delta(\boldsymbol{D})} = (\delta(d_{mi}))_{1 \leq m \leq M, 1 \leq i \leq I}.
\end{equation}
We can see the matrix $\boldsymbol{\delta(\boldsymbol{D})}$ as a denoising version of $\boldsymbol{D}$, i.e, the shrinkage rule $\delta(d)$ acts by denoising the empirical coefficient $d$ in order to estimate $\theta$ in \eqref{coefmodel}, $\delta(d) = \hat{\theta}$. Thus, the estimation $\boldsymbol{\hat{\Gamma}}$ of the wavelet coefficients matrix $\boldsymbol{\Gamma}$ is given by least squares,
\begin{equation}
\boldsymbol{\hat{\Gamma}} = \boldsymbol{\delta(\boldsymbol{D})}\boldsymbol{y^{t}}\boldsymbol{(yy^{t})^{-1}},
\end{equation}
and finally $\boldsymbol{\alpha}$ can be estimated at locations $t_1, \cdots, t_M$ by the inverse discrete wavelet transformation (IDWT),
\begin{equation}
\boldsymbol{\hat{\alpha}} = \boldsymbol{W^t}\boldsymbol{\hat{\Gamma}}.
\end{equation}
There are several wavelet shrinkage rules available in the literature and the readers are adressed to Vidakovic (1999) for descriptions of most of them. In this work, we consider the bayesian wavelet shrinkage rule proposed by Sousa (2020) that assumes a mixture of a point mass function at zero and a logistic distribution as prior distribution of a single linear combination of wavelet coefficients $\theta$,
\begin{equation}\label{prior}
\pi(\theta;p,\tau) = p \delta_{0}(\theta) + (1-p)g(\theta;\tau),
\end{equation}
where $p \in (0,1)$, $\delta_{0}(\theta)$ is the point mass function at zero and $g(\theta;\tau)$ is the logistic density function symmetric around zero, for $\tau > 0$, 
\begin{equation} \label{log}
g(\theta;\tau) = \frac{\exp\{-\frac{\theta}{\tau}\}}{\tau(1+\exp\{-\frac{\theta}{\tau}\})^2}\mathbb{I}_{\mathbb{R}}(\theta).
\end{equation} 
Under squared loss function, the associated bayesian shrinkage rule is the posterior expected value of $\theta$, $\mathbb{E}_{\pi}(\theta|d)$, that under the model prior \eqref{prior}, is given by (Sousa, 2020),
\begin{equation}\label{rule}
\delta(d) = \mathbb{E}_{\pi}(\theta|d) = \frac{(1-p)\int_\mathbb{R}(\sigma u + d)g(\sigma u +d ; \tau)\phi(u)du}{\frac{p}{\sigma}\phi(\frac{d}{\sigma})+(1-p)\int_\mathbb{R}g(\sigma u +d ; \tau)\phi(u)du},
\end{equation}
where $\phi(\cdot)$ is the standard normal density function. The shrinkage rule \eqref{rule} under the model \eqref{log} is called logistic shrinkage rule and has interesting features under estimation point of view. First, its hyperparameters $p$ and $\tau$ control the degree of shrinkage of the rule. Higher values of $\tau$ or $p$ imply higuer shrinkage level, i.e, the rule will reduce severely the magnitudes of the empirical coefficients. Further, as described in Sousa (2020), the logistic shrinkage rule had good performances in terms of averaged mean squared error in simulation studies against standard shrinkage or thresholding procedures. Figure \ref{fig:rules} (a) shows the logistic shrinkage rule obtained numerically for some values of $\tau$ in the interval $(0.5,50)$, $p = 0.9$ and $\sigma = 1$. 

\begin{figure}[H]
\centering
\subfigure[Bumps function.]{
\includegraphics[scale=0.5]{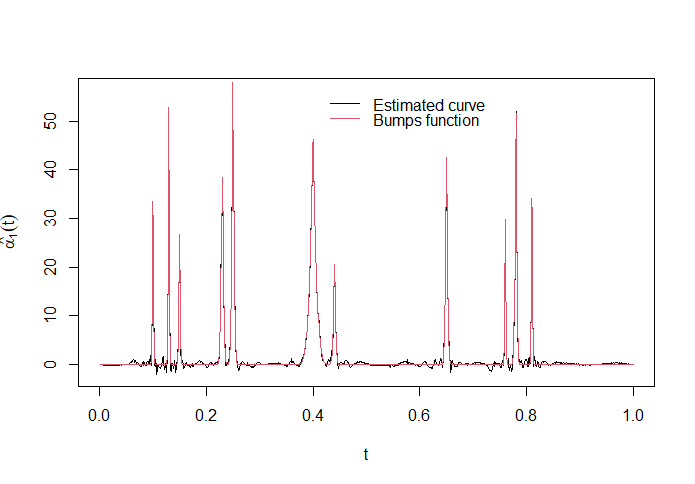}}
\subfigure[Blocks function.]{
\includegraphics[scale=0.5]{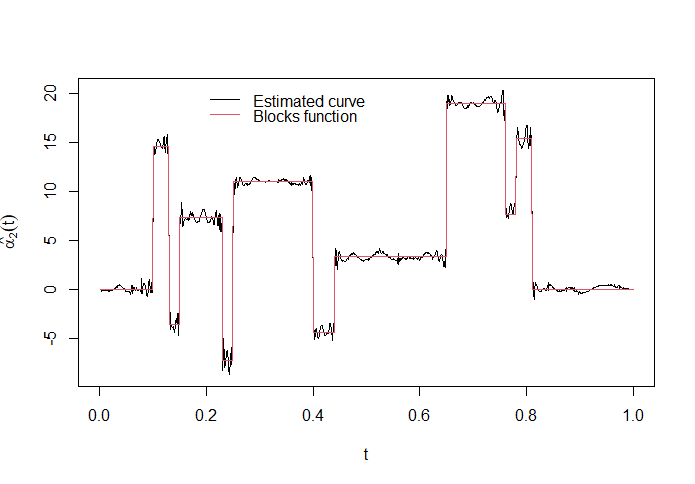}}
\caption{Bumps and blocks estimated functions by the logistic wavelet shrinkage rule with hyperparameters $p = 0.9$ and $\tau = 5$.} \label{fig:ex1c}
\end{figure}

Figures \ref{fig:ex1c} (a) and (b) show the estimated Bumps and Blocks functions respectively by the application of the logistic wavelet shrinkage rule to the generated aggregated datasets of Figure \ref{fig:ex1b} (a) and described in the Section 2. We chose the hyperparameters values $p = 0.9$ and $\tau = 5$ in \eqref{rule} to perform the estimation. Further, the DWT was applied to the datasets by considering Daubechies basis function with ten null moments ($N=10$), which wavelet function $\psi$ is shown in Figure \ref{fig:rules} (b),  and the standard deviation $\sigma$ was estimated  according to Donoho and Johnstone (1994) proposal,
\begin{equation}\label{eq:sigma}
\hat{\sigma} = \frac{\mbox{median}\{|d_{J-1,k}|:k=0,...,2^{J-1}\}}{0.6745}.
\end{equation}
It is possible to note in the figures that local features of the functions were estimated, such as the discontinuities of the Blocks function and spikes of the Bumps function. Excessive values of $p$ and/or $\tau$ can underestimate the magnitude of these local features by reducing empirical coefficients magnitudes too much. Then, a suitable hyperparameters elicitation should allow the rule to reduce coefficients magnitudes but without loosing the ability to estimate important local characteristics of the functions to be recovered. A very useful proposal of elicitation of the hyperparameter $p$ is done by Angelini and Vidakovic (2004), 
\begin{equation}\label{eq:alpha}
p = p(j) = 1 - \frac{1}{(j-J_{0}+1)^2},
\end{equation}
where $J_ 0 \leq j \leq J-1$, $J_0$ is the primary resolution level. For $\tau$ elicitation, Sousa (2020) suggests $\tau \in \{1,\cdots, 10\}$. 

\begin{figure}[H]
\centering
\subfigure[Aggregated curves.]{
\includegraphics[scale=0.35]{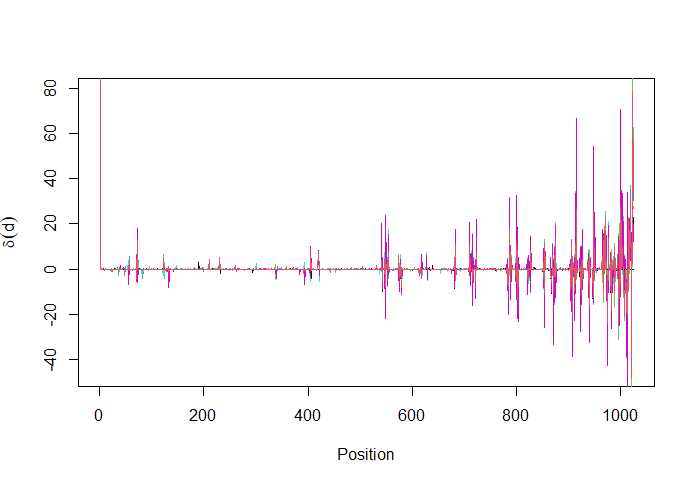}}
\subfigure[Bumps function.]{
\includegraphics[scale=0.35]{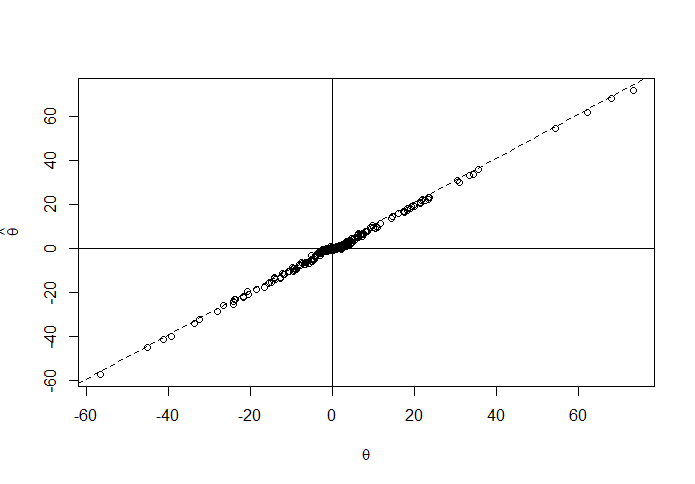}}
\subfigure[Blocks function.]{
\includegraphics[scale=0.35]{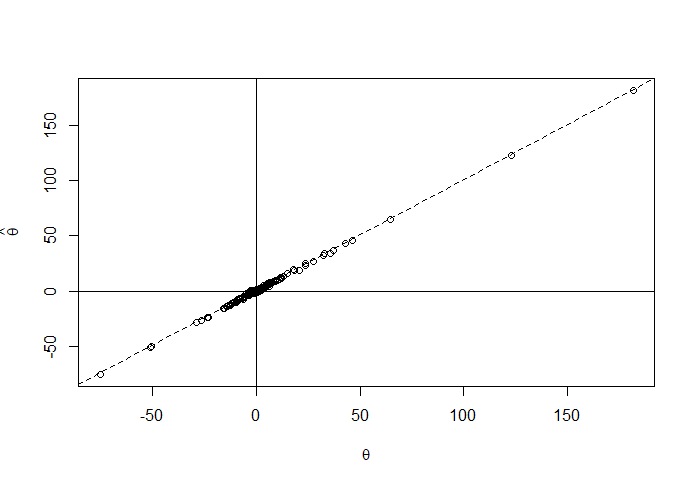}}
\caption{Estimated wavelet coefficients of the aggregated curves (a), estimated wavelet coefficients against their true values for Bumps (b) and Blocks (c) functions (dashed lines $y=x$).} \label{fig:ex1d}
\end{figure}

Figure \ref{fig:ex1d} (a) presents the 1024 estimated (shrunk) wavelet coefficients of the aggregated curves ($\boldsymbol{\theta}$) and Figures \ref{fig:ex1d} (b) and (c) show the estimated wavelet oefficients of the component functions Bumps ($\gamma_{m1}$, $m=1,\cdots,1024$) and Blocks ($\gamma_{m2}$, $m=1,\cdots,1024$) against their true values respectively. In fact, we observe the shrinkage rule action by the reduced magnitudes of the estimated wavelet coefficients in comparison with the empirical ones. Sufficiently small empirical coefficients were set to zero or very close to zero by the rule. Only significant coefficients were preserved in terms of magnitudes. Moreover, the estimated coefficients were very close to the true values for both Bumps and Blocks functions, once the points $(\theta, \hat{\theta})$ are most of them very close to the line $y = x$ (dashed line in Figures \ref{fig:ex1d} (b) and (c)). 

\section{Monte Carlo Simulation Studies}

\begin{figure}[H]
\centering
\subfigure[Doppler.]{
\includegraphics[scale=0.35]{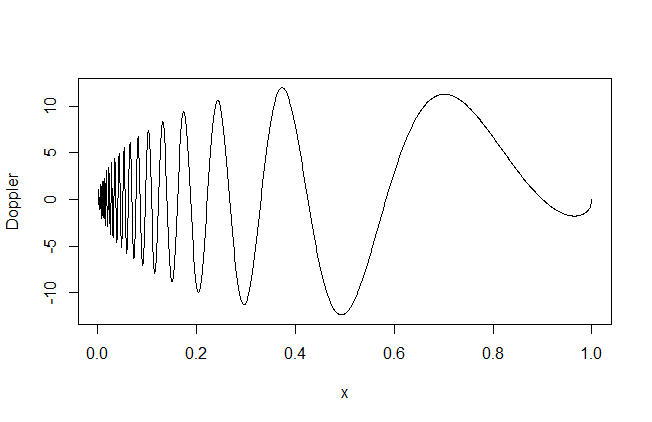}}
\subfigure[Heavisine.]{
\includegraphics[scale=0.35]{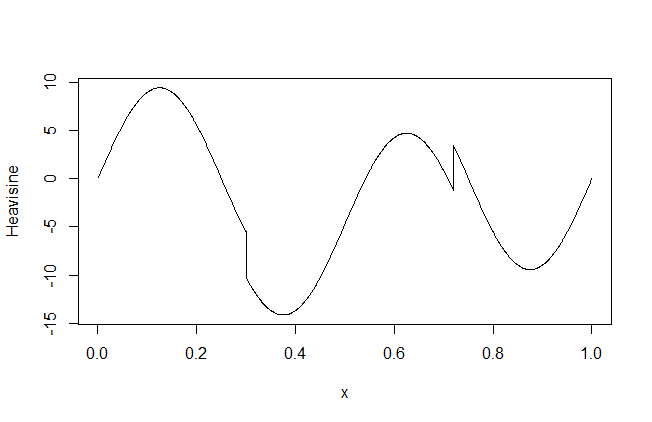}}
\subfigure[Logit.]{
\includegraphics[scale=0.35]{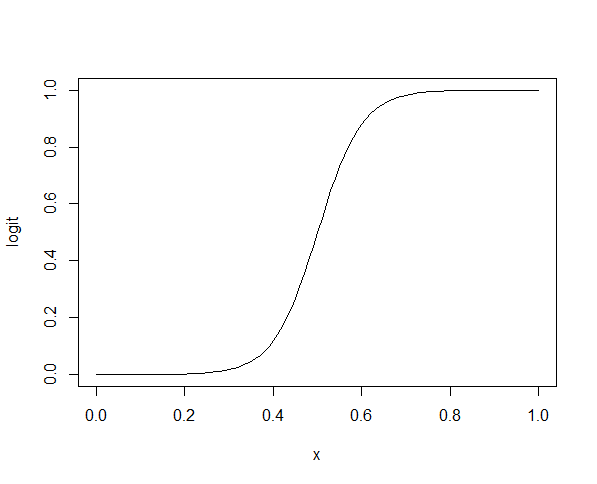}}
\subfigure[SpaHet.]{
\includegraphics[scale=0.35]{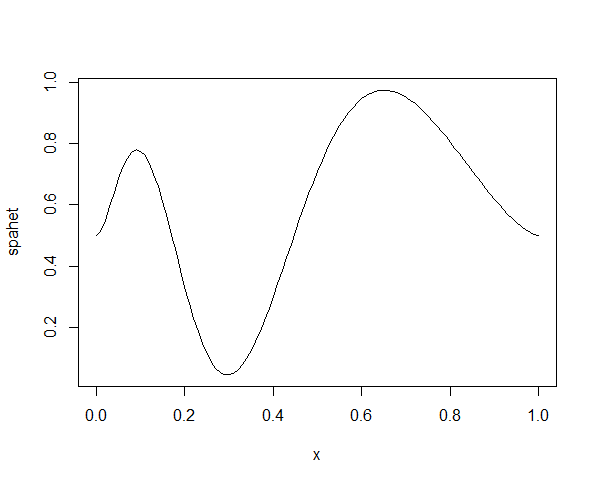}}
\caption{Doppler (a) and Heavisine (b) Donoho-Johnstone test functions, logit (c) and spatially heterogeneous (spaHet) (d) functions used as component functions in the simulation studies.} \label{fig:functions}
\end{figure}

We conducted three Monte Carlo simulation studies to evaluate the performance of the proposed wavelet-based method. The first study considered the two component functions ($L = 2$) of the previous example, Bumps and Blocks functions. The second one took the four Donoho and Johnstone test functions as component functions ($L = 4$), Bumps, Blocks, Doppler and Heavisine functions. Finally, the third simulation study considered two smooth functions called Logit and Spatially Heterogeneous (SpaHet) by Wand (2000), Ruppert (2002) and Goepp et al. (2018), as component functions ($L = 2$). The definitions of the functions are given in Table \ref{tab:compfun} and the plots of Doppler, Heavisine, Logit and SpaHet in the interval $[0,1]$ are in Figure \ref{fig:functions}.

Thus, the first two simulation studies have the goal of evaluating the proposed method in the cases where the component functions have local characteristics to be recovered. In addition to the already discussed Bumps and Blocks functions, Doppler function has oscillation behaviour and Heavisine function has two discontinuities at $t = 0.3$ and $0.72$. The third simulation study was conducted to evaluate how the method works when the component functions are completely smooth, which are the case of Logit and SpaHet functions.

\begin{figure}[H]
\centering
\includegraphics[scale=0.38]{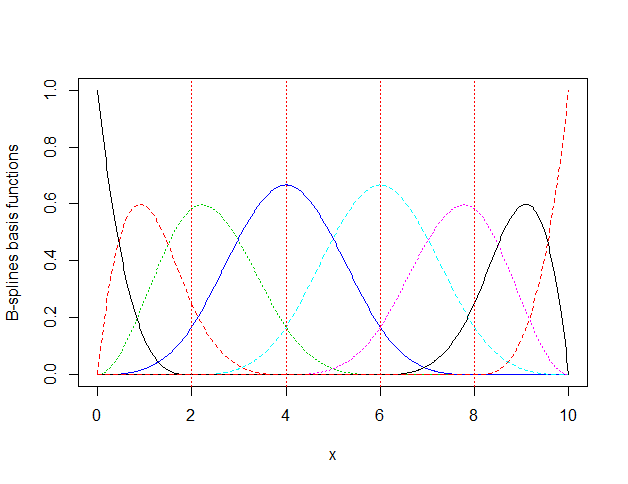}
\caption{Eight cubic B-splines basis defined by four interior knots.}\label{fig:bsplines}
\end{figure}

The performances of the wavelet-based method in the simulation studies are compared with a B-spline basis expansion of the component functions, which are more adapted for computational implementation than splines basis due to its compact support. The $i$-th B-spline of order $m$ and knots $a=t_0 < t_1 <\cdots<t_{K}<t_{K+1}=b$, denoted by $B_{i,m}(x)$ is a spline of order $m$ defined on the same knots such that is nonzero over at most $m$ consecutive subintervals, i.e, over $[t_i,t_{i+m}]$ and zero outside it and can be defined recursively by
\begin{equation}
B_{i,m}(x) = \frac{x - t_i}{t_{i+m} - t_i}B_{i,m-1}(x) + \frac{t_{i+m+1} - x}{t_{i+m+1}-t_{i+1}}B_{i+1,m-1}(x),
\end{equation}    
where $B_{i,1}(x) = \mathbb{I}_{[t_i,t_{i+1})}(x)$. Figure \ref{fig:bsplines} shows eight cubic B-splines defined by four interior and equally spaced knots in $[0,10]$. For more details about B-splines, see De Boor (1978) and Ramsay and Silverman (2005).

\begin{table}[H]
\scalefont{0.80}
\centering
\label{my-label}
\begin{tabular}{|c|}
\hline
\textbf{BUMPS} \\ 
$ f(x) = \sum_{l=1}^{11} h_l K\left(\frac{x - x_l}{w_l} \right)$ \\
where \\

$K(x) = (1 + |x|)^{-4}$ \\

$(x_l)_{l=1}^{11} = (0.1, 0.13, 0.15, 0.23, 0.25, 0.40, 0.44, 0.65, 0.76, 0.78, 0.81)$\\

$(h_l)_{l=1}^{11} = (4, 5, 3, 4, 5, 4.2, 2.1, 4.3, 3.1, 5.1, 4.2)$  \\

$(w_l)_{l=1}^{11} = (0.005, 0.005, 0.006, 0.01, 0.01, 0.03, 0.01, 0.01, 0.005, 0.008, 0.005)$\\ \hline 

\textbf{BLOCKS} \\
$ f(x) = \sum_{l=1}^{11} h_l K(x - x_l)$ \\
where \\

$K(x) = (1 + \mathrm{sgn}(x))/2$ \\

$(x_l)_{l=1}^{11} = (0.1, 0.13, 0.15, 0.23, 0.25, 0.40, 0.44, 0.65, 0.76, 0.78, 0.81)$  \\

$(h_l)_{l=1}^{11} = (4, -5, 3, -4, 5, -4.2, 2.1, 4.3, -3.1, 2.1, -4.2)$ \\ \hline 

\textbf{DOPPLER} \\
$ f(x) = \sqrt{x(1-x)}\sin\left(\frac{2.1 \pi}{x + 0.05} \right)$ \\ \hline 

\textbf{HEAVISINE} \\
$ f(x) = 4\sin(4 \pi x) - \mathrm{sgn}(x - 0.3) - \mathrm{sgn}(0.72 - x)$ \\ \hline

\textbf{LOGIT} \\ 
$f(x) = \frac{1}{1+\exp\{-20(x-0.5)\}} $ \\ \hline

\textbf{SPAHET}  \\ 
$f(x) = \sqrt{x(1-x)}\sin\left(\frac{2\pi(1+2^{-0.6})}{x+2^{-0.6}}\right)$ \\ \hline

\end{tabular}
\caption{Test functions definitions for $x \in [0,1]$ used as component functions in the simulation studies.}\label{tab:compfun}
\end{table}

For each simulation study, $I = 50$ samples were generated according to model \eqref{discmodel}, considering $M = 512$ and $1024$ data points for each sample. Further, the random noises were generated with $\sigma^2$ according to two signal to noise ratio values, $\mathrm{SNR} = 3$ and $9$. For each scenario of $M$ and $\mathrm{SNR}$, $N = 100$ replicates were done and the component functions points were estimated according to the proposed method. The mean squared error (MSE) of the j-the replication of a component function $\alpha_l(t)$, 
\begin{equation}
\mathrm{MSE_l^{(j)}} = \frac{1}{M} \sum_{i=1}^{M}[{\hat \alpha_l^{(j)}(t_i)} - \alpha_l(t_i)]^2, \nonumber
\end{equation}
was calculated for each replicate and the averaged mean squared error (AMSE), 
\begin{equation}
\mathrm{AMSE_l} = \frac{1}{N} \sum_{j=1}^{N}\mathrm{MSE}_l^{(j)}, \nonumber
\end{equation}
was considered as performance measure $l = 1,\cdots,L$. 

The results are available in Tables \ref{tab:bb}, \ref{tab:bbdh} and \ref{tab:ls} for aggregated data with Bumps and Blocks (simulation study 1), Bumps, Blocks, Doppler and Heavisine (simulation study 2) and Logit and SpaHet (simulation study 3) respectively. In general, the proposed wavelet-based method outperformed the B-spline-based method in almost all the scenarios of simulation studies 1 and 2, as expected, since the component functions have local features such as discontinuities, spikes and oscillations that are well recovered by wavelet basis expansions but are not by splines. In simulation study 3, the  B-spline expansion had better results in terms of AMSE than by wavelets, once the component functions are smooth in their whole domains. It is important to note that although the estimation by splines were better in this last simulation study, the wavelet-based estimation method had great work and estimated the component functions Logit and SpaHet well in all the scenarios of this study. 

Looking only at wavelets-based method results, no meaningful improvement in AMSE was observed according to the sample size $n = 1024$ against $n = 512$, but better results were observed for $\mathrm{SNR} = 9$ against $\mathrm{SNR = 3}$ as also expected.

\begin{table}[H]
\scalefont{0.7}
\centering
\label{my-label}
\begin{tabular}{|c|c|c|c|c|}
\hline
 \multicolumn{5}{|c|}{\textbf{Simulation study 1}} \\ \hline
\textbf{n} & \textbf{Function}  & \textbf{Method} & \textbf{SNR = 3} & \textbf{SNR = 9}  \\ \hline \hline
512          & Bumps & Wavelets & \textbf{0.2942 (0.0195)} & \textbf{0.3312 (0.0107)}\\ 
               &     & Splines & 31.2887 (0.0063) &  31.2621 (0.0006) \\ \hline
         & Blocks    & Wavelets & \textbf{0.2680 (0.0184)} &  \textbf{0.2637 (0.0082)} \\
               &     & Splines  & 4.8235 (0.0059) & 4.7968 (0.0006) \\ \hline \hline  
1024         & Bumps & Wavelets & \textbf{0.2462 (0.0127)} & \textbf{0.2182 (0.0052)} \\
               &      & Splines & 31.2604 (0.0031) & 31.2467 (0.0003) \\ \hline
          & Blocks & Wavelets & \textbf{0.2222 (0.0128)} & \textbf{0.1788 (0.0052)} \\
               &      & Splines & 5.0421 (0.0034) & 5.0282 (0.0003) \\ \hline

\end{tabular}
\caption{AMSE (standard deviation) of simulation study 1 for aggregated data generated with the component functions Bumps and Blocks.}\label{tab:bb}
\end{table}

\begin{table}[H]
\scalefont{0.7}
\centering
\label{my-label}
\begin{tabular}{|c|c|c|c|c|}
\hline
 \multicolumn{5}{|c|}{\textbf{Simulation study 2}} \\ \hline
\textbf{n} & \textbf{Function}  & \textbf{Method} & \textbf{SNR = 3} & \textbf{SNR = 9}  \\ \hline \hline
512          & Bumps & Wavelets & \textbf{0.3721 (0.0271)} & \textbf{0.4064 (0.0122)}\\ 
               &     & Splines & 31.2990 (0.0081) & 31.2633 (0.0009) \\ \hline
         & Blocks    & Wavelets & \textbf{0.4698 (0.0358)} & \textbf{0.3637 (0.0162)}  \\
               &     & Splines  & 4.8553 (0.0127) & 4.8003 (0.0012) \\ \hline  
          & Doppler    & Wavelets & \textbf{1.5954 (0.1279)} & \textbf{0.7020 (0.0415)} \\
               &     & Splines  & 3.1705 (0.0555) & 2.9021 (0.0072) \\ \hline 
         & Heavisine    & Wavelets & 1.3009 (0.1119) & 0.5456 (0.0313) \\
               &     & Splines  & \textbf{0.3354 (0.0479)} & \textbf{0.1335 (0.0052)} \\ \hline \hline
1024         & Bumps & Wavelets & \textbf{0.3042 (0.0156)} & \textbf{0.2703 (0.0070)} \\
               &      & Splines & 31.2654 (0.0042) & 31.2474 (0.0004)  \\ \hline
          & Blocks & Wavelets & \textbf{0.3837 (0.0200)} & \textbf{0.2315 (0.0088)} \\
               &      & Splines & 5.0567 (0.0063) & 5.0299 (0.0007)  \\ \hline   
         & Doppler    & Wavelets & \textbf{1.2996 (0.0835)} & \textbf{0.4466 (0.0277)} \\
               &     & Splines  & 3.1923 (0.0313) & 3.0571 (0.0034) \\ \hline                                      
         & Heavisine    & Wavelets & 1.0270 (0.0634) & 0.3870 (0.0229)\\
              &     & Splines  & \textbf{0.2218 (0.02544)} & \textbf{0.1221 (0.0028)} \\ \hline
  
\end{tabular}
\caption{AMSE (standard deviation) of simulation study 2 for aggregated data generated with the component functions Bumps, Blocks, Doppler and Heavisine.}\label{tab:bbdh}
\end{table} 

\begin{table}[H]
\scalefont{0.7}
\centering
\label{my-label}
\begin{tabular}{|c|c|c|c|c|}
\hline
 \multicolumn{5}{|c|}{\textbf{Simulation study 3}} \\ \hline
\textbf{n} & \textbf{Function}  & \textbf{Method} & \textbf{SNR = 3} & \textbf{SNR = 9}  \\ \hline \hline
512          & Logit & Wavelets & 0.0267 (0.0001) & 0.0267 ($4 \times 10^{-5}$) \\ 
               &     & Splines & \textbf{0.0001 ($\boldsymbol{2 \times 10^{-5}}$)} & $\boldsymbol{1 \times 10^{-5}}$ ($\boldsymbol{2 \times 10^{-6}}$)\\ \hline
         & SpaHet    & Wavelets & 0.0217 (0.0004) & 0.0217 (0.0001) \\
               &     & Splines  & \textbf{0.0001} ($\boldsymbol{2 \times 10^{-5}}$) & $\boldsymbol{1 \times 10^{-5}}$ ($\boldsymbol{2 \times 10^{-6}}$)\\ \hline   \hline
1024         & Logit & Wavelets & 0.0221 (0.0001) & 0.0222 ($4 \times 10^{-5}$) \\
               &      & Splines & $\boldsymbol{5 \times 10^{-5}}$ ($\boldsymbol{1 \times 10^{-5}}$) & $\boldsymbol{6 \times 10^{-6}}$ ($\boldsymbol{6 \times 10^{-6}}$) \\ \hline
          & SpaHet & Wavelets & 0.0126 (0.0001) & 0.0127 ($6 \times 10^{-5}$)\\
               &      & Splines & $\boldsymbol{5 \times 10^{-5}}$ ($\boldsymbol{1 \times 10^{-5}}$) & $\boldsymbol{5 \times 10^{-6}}$ ($\boldsymbol{1 \times 10^{-6}}$) \\ \hline

\end{tabular}
\caption{AMSE (standard deviation) of simulation study 3 for aggregated data generated with the component functions Logit and SpaHet.}\label{tab:ls}
\end{table}

\section{Tecator Dataset Application}
\begin{figure}[H]
\centering
\subfigure[Tecator dataset.]{
\includegraphics[scale=0.55]{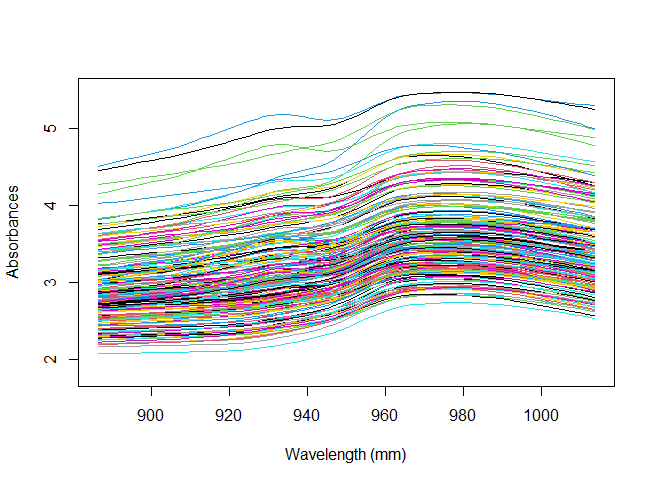}}
\subfigure[Empirical wavelet coefficients.]{
\includegraphics[scale=0.55]{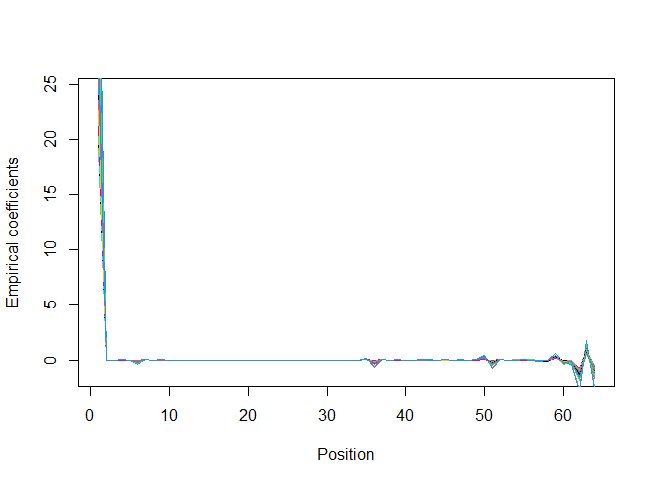}}

\caption{Tecator dataset (a) and the empirical wavelet coefficients of the curves (b).} \label{fig:tecator1}
\end{figure}

We applied the proposed wavelet-based procedure in the so called Tecator dataset, that involves absorbance curves of $I = 215$ meat samples in $M = 64$ equally spaced wavelength points from 850 to 1050 mm. Each meat sample can be decomposed by fat, water and protein components with different concentrations among the samples. The dataset was obtained by the R package fda.usc by Bande and de la Fuente (2012) and is shown in Figure \ref{fig:tecator1} (a). The empirical wavelet coefficients of the samples obtained by the DWT application with Daubechies basis (10 null moments) are in Figure \ref{fig:tecator1} (b). Note that the aggregated curves are smooth along the domain and most of the empirical wavelet coefficients are zero or very close to zero.

\begin{figure}[H]
\centering
\subfigure[Fat curve.]{
\includegraphics[scale=0.35]{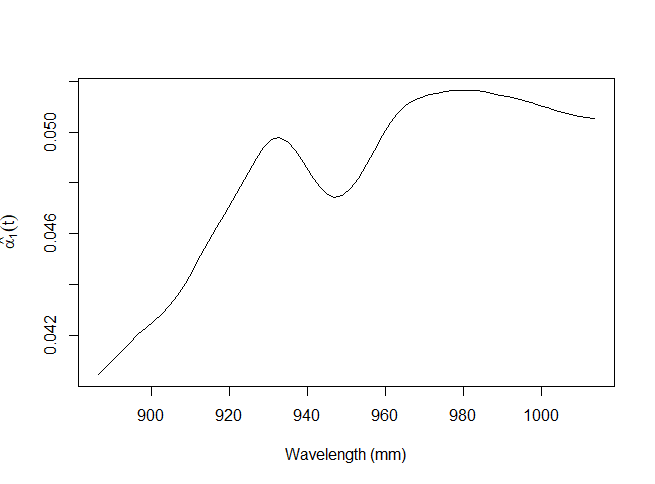}}
\subfigure[Water curve.]{
\includegraphics[scale=0.35]{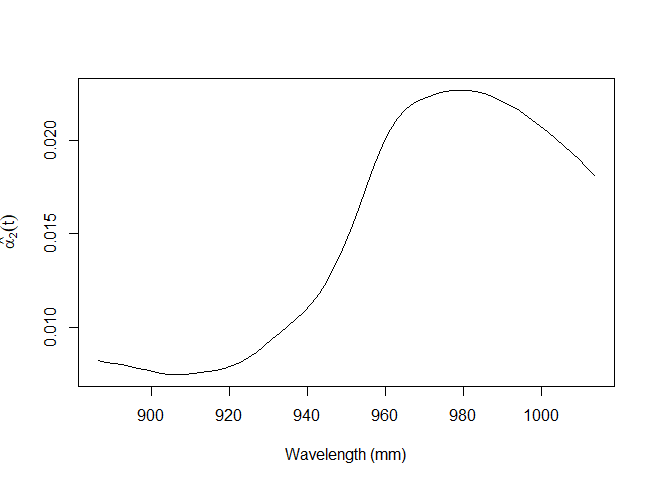}}
\subfigure[Protein curve.]{
\includegraphics[scale=0.35]{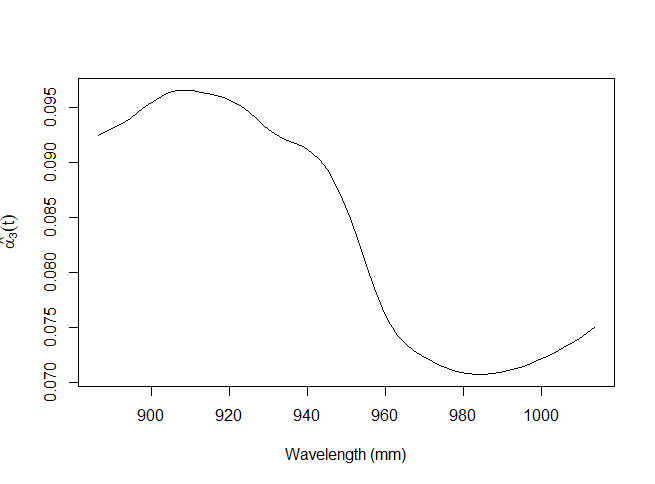}}
\caption{Estimated fat (a), water (b) and protein (c) curves by application of logistic wavelet shrinkage rule.} \label{fig:tecator2}
\end{figure}

According to Beer-Lambert Law (Brereton, 2003), we can write the absorbance $A(t)$ of a meat sample at wavelength $t$ as a linear combination of the absorbances $\alpha_1(t)$, $\alpha_2(t)$ and $\alpha_3(t)$ of its components, fat, water and protein respectively, as model \eqref{funmodel}, 
\begin{equation}\label{funmodeltec}
A(t) = \sum_{l=1}^{3}y_{l} \alpha_{l}(t) + e(t),
\end{equation}
where the weights $y_l$ are the concentrations of the components in the meat sample. Figures \ref{fig:tecator2}  (a), (b) and (c) show the estimated fat, water and protein curves respectively by the application of logistic shrinkage rule with hyperparameters $\tau = 5$ and level dependent $\alpha$ values according to the \eqref{eq:alpha}. In fact, the component curves have different features. The water curve is increasing in the range of 900 to 980 mm and reaches a maximum point at approximately 980 mm. The protein curve is decreasing along the same interval, reaching a minimum point at 980 mm. Finally, the fat curve has two peaks at 930 and 980 mm.

\section{Conclusion and Further Considerations}
This work proposed the use of wavelet basis to estimate individual mean curves from samples of aggregated curves composed of linear combinations of these curves. Furthermore, it is proposed the application of the bayesian shrinkage rule associated with a prior mixture of a Dirac function at zero and the logistic distribution symmetric at zero.

In fact, the proposed methodology considers the functional structure of the data, in addition to allowing the estimation of local characteristics of the individual curves, such as discontinuities, peaks and oscillations. It occurs due to the fact that the wavelets are well located in time domain, which is not the case of other function basis, such as splines or Fourier. Multivariate methods also present problems to estimate curves with such profiles. Regarding the bayesian shrinkage rule, the hyperparameters allow controlling the degree of shrinkage of the estimator, even allowing them to be dependent on the resolution level.
The simulation studies showed excellent performance of the proposed methodology in terms of AMSE in comparison with B- splines based method, with better performance against this last one when individual curves present discontinuities, peaks or oscillations.

The study of the performance of the proposal for different shrinkage rules (Bayesian or not) and the use of different wavelet family basis are of interest for future studies.

\section*{Acknowledgements}

This article was supported by CAPES\footnote{Coordination of Superior Level Staff Improvement, Brazil} fellowship.

\end{document}